\documentclass[5p]{elsarticle}

\usepackage{amsfonts,amsmath,amssymb,amsthm,bbm,hyperref}
\usepackage{graphicx}

\newcommand{\be}{\begin{equation}}
\newcommand{\ee}{\end{equation}}
\newcommand{\beq}{\begin{eqnarray}}
\newcommand{\eeq}{\end{eqnarray}}
\newcommand{\nn}{\nonumber}

\newcommand{\f}{\frac}
\newcommand{\vphi}{\varphi}

\newcommand{\Christoffel}[1]{{\,\!}^{#1}\Gamma}

\begin{document}

\title{Classical geodesics from the canonical quantisation of  spacetime coordinates}

\author[1]{I\~naki Garay}
\ead{inaki.garay@ehu.eus}

\author[2,3]{Salvador Robles-P\'{e}rez}
\ead{salvador.robles@educa.madrid.org}

\address[1]{Fisika Teorikoa eta Zientziaren Historia Saila, UPV/EHU,
  644 P.K., 48080 Bilbao, Spain}
  
\address[2]{Departamento de matem\'{a}ticas, IES Miguel Delibes, Miguel Hern\'{a}ndez 2, 28991 Torrej\'{o}n de la Calzada, Spain}  

\address[3]{Estaci\'{o}n Ecol\'{o}gica de Biocosmolog\'{\i}a, Pedro de Alvarado, 14, 06411 Medell\'{\i}n, Spain}

\begin{abstract}
A canonical quantisation of the coordinates of the spacetime within the general relativity theory is proposed. This quantisation will depend on the observer but it provides an interesting perspective on the problem of relating the non-relativistic and classical limits of a possible quantum gravity theory. In this sense, within  this formalism, it is possible to recover from the quantum equation satisfied by a test field the classical geodesics of the corresponding spacetime. On the other hand, the Schr\"odinger equation is recovered in the Newtonian limit. A key ingredient for this procedure is a generalization of the Maupertuis principle, that is, the possibility of describing the geodesics of a given metric as the non-affine geodesics of other conformal metric under the action of a potential.
\end{abstract}

\begin{keyword}
Canonical quantisation \sep geodesics \sep Newtonian limit

\PACS 04.20 \sep 03.65.-w \sep 45.20.D-

\end{keyword}

\maketitle


\section{Introduction}

The path to a full understanding of the gravitational interaction at the quantum level is still unclear. Despite of the many advances that have taken place in this direction during the last three decades there are deep issues that remain unsolved \cite{Kiefer:2013jqa,Strominger:2009aj}. 
In this article we treat this problem from the perspective of the canonical quantisation of the spacetime coordinates. 
This point of view provides new tools to study the problem and offers interesting relations between the different limits of a possible quantum gravity theory, i.e., the non-relativistic and the classical limit.

In order to compute the geodesics of the spacetime given by a certain metric that is a solution to the Einstein equations we will consider an action principle, invariant under reparametrizations \cite{Blau2016}, and we will apply the canonical quantisation procedure to the spacetime coordinates. In principle, nothing new is expected. Nevertheless, we shall present three interesting results that appear as a consequence of it.

First, we shall make use of an extension of the Maupertuis principle \cite{Biesiada1994} to show that the trajectory followed by a test particle that moves in a curved spacetime under the action of a given potential can be seen as the geodesic followed by the particle in a conformal spacetime.
Reversely, the geodesic followed by a particle in a curved spacetime can be seen as the trajectory of the test particle moving in a curved space under the action of a certain potential. In particular, applying the formalism to the static coordinates of de Sitter spacetime we show that the geodesic followed by a test particle can be seen as the trajectory followed by the particle in a related curved space under the action of a central potential. As  expected, the test particle, which could be for instance a distant galaxy, moves in an accelerated way until it approaches the cosmological horizon, where its velocity becomes frozen.

On the other hand, it is well known that the canonical quantisation of the Hamiltonian constraint gives rise to the Klein-Gordon equation (see, for instance, Ref. \cite{Chernikov1968}). 
However, this procedure, unlike the customary one based on the variation of the action associated to a classical scalar field, introduces the Planck constant in the solution of the Klein-Gordon equation. 
It allows us to expand the scalar field in  powers of $\hbar$. Following then a procedure that is extensively used in quantum cosmology to obtain the semiclassical expansions of the wave function (see, for instance, Refs. \cite{Halliwell1990, Wiltshire2003, Kiefer2007}), we obtain at zeroth order in $\hbar$ the equation of the geodesic of the spacetime where the field propagates. Therefore, we show that the solution of the Klein-Gordon equation contains not only information about the matter that propagates in the spacetime but also information about the geometrical structure of the spacetime itself through the equation of the geodesics. 

Finally, we consider the Newtonian limit of the curved spacetime to show that, in that limit, the Klein-Gordon equation can be written as the Schr\"{o}dinger equation for a particle that moves in a nearly flat space under the action of a given potential. Thus, the canonical quantisation of the spacetime coordinates relates the classical description of a curved spacetime in the theory of general relativity with the quantum description of a field in a quantum field theory as well as their non relativistic limits, the Newtonian limit of classical mechanics and the Schr\"{o}dinger equation of quantum mechanics.

The paper is outlined as follows. We will explore in section \ref{Lagrangian} the consequences of the Maupertius principle applied to the case of the geodesics of the spacetime. More concretely, in section \ref{sec2A} we will develop the Lagrangian and the Hamiltonian formalism for the action that will be used to obtain the geodesics. 
Furthermore, in section \ref{Newtoniansection} we will explore the Newtonian limit making use of a conformal metric and the potential associated to it and,  in section \ref{deSitter}, we will apply our results to the case of the de Sitter spacetime in static coordinates.

The quantisation of the spacetime coordinates and its consequences will be treated in section \ref{Quant}. We will review the prescription to canonically quantise the spacetime coordinates and to obtain the Klein-Gordon equation. Afterwards, in section \ref{geodesic} we will explore how to recover the classical geodesic equation taking as starting point the Klein-Gordon equation associated with the quantised coordinates of the spacetime. 
Furthermore, the explicit computation of the geodesics of de Sitter universe from the state given by the Bunch-Davies vacuum will be performed in section \ref{BDsection}. Finally, in section \ref{Sch}, we will study the recovery of the Schr\"odinger equation in the appropriate limit.


\section{Geodesics of the spacetime}\label{Lagrangian}

As it is well known \cite{Misner1970, Wald1984}, the equation of the geodesic of a given spacetime with metric tensor $g_{\mu\nu}$ can be obtained from the variational principle of the action
\be\nn
S = \int \sqrt{g_{\mu\nu} \dot x^\mu \dot x^\nu} d\tau ,
\ee
where $\dot x^\mu \equiv \frac{dx^\mu}{d\tau}$, and $\tau$ is an affine parameter, i.e. $ds = m d\tau$, with $m$ being a constant. The action $S$ is invariant under reparametrization of the geodesics. 
However, the presence of the square root in the Lagrangian makes the momenta conjugated to the spacetime coordinates to be non linear functions whose quantisation becomes  cumbersome. 
For this purpose, it is more useful to consider the action
\be
\tilde S = \frac{1}{2} \int \left( g_{\mu\nu} \dot x^\mu \dot x^\nu \right) d\tau ,\label{tildeS}
\ee
whose variational principle gives rise to the same geodesic equation \cite{Misner1970, Wald1984, Hartle1976, Bertschinger1999, Blau2016} and the corresponding Lagrangian is quadratic in the velocities (i.e. the tangent vectors).

In this section we will use a slightly modified version of equation \eqref{tildeS} that is invariant under reparametrizations \cite{Blau2016} (notice that \eqref{tildeS} is not). We will explore the possibility of describing the geodesic equation with other conformal metric under a potential and we will study the Newtonian limit of the theory. Finally, an application to the example of the de Sitter universe will be studied.

\subsection{Lagrangian and Hamiltonian formulation of spacetime geodesics}\label{sec2A}

Let us consider the following action as the starting point for the Lagrangian formulation of the geodesics:
\beq\label{ACT01}
\hspace*{-0.5cm}S[x^\mu(\tau)] &=& \int L(x,\dot{x}) d\tau\nonumber\\
& =& \!\frac{m}{2}\!\!\int\!\!\left(\!\frac{1}{N^2} g_{\mu\nu} \dot{x}^\mu \dot{x}^\nu\!- (1+V(x)) \!\right)\! N d\tau ,\label{A01}
\eeq
where, $\dot{f} \equiv \frac{d f}{d \tau}$, $\tau$ is the parameter of the curve, $V(x)$ a potential and $g_{\mu\nu}$ the metric tensor of the $3+1$ dimensional spacetime with line element given by
\be
ds^2 = g_{\mu\nu} dx^\mu dx^\nu ,
\ee
where we have chosen the signature $(-,+,+,+)$. Note that the action is invariant under reparametrizations $Nd\tau=\tilde{N}d\tilde{\tau}$.

The variational principle of the action integral (\ref{A01}) with respect to $x^\mu(\tau)$ gives rise to the following geodesic equation
\be\label{GEO01}
\ddot{x}^\mu + \Christoffel{g}^\mu_{\alpha \beta} \dot{x}^\alpha \dot{x}^\beta = \frac{\dot{N}}{N}\dot{x}^\mu - \frac{N^2}{2}g^{\mu\nu}\partial_\nu V,
\ee
with
\be\label{Chr01}
\Christoffel{g}^\mu_{\alpha \beta} = \frac{1}{2} g^{\mu\nu} \left( \frac{\partial g_{\alpha\nu}}{\partial x^\beta} + \frac{\partial g_{\beta\nu}}{\partial x^\alpha} -  \frac{\partial g_{\alpha\beta}}{\partial x^\nu} \right)\,.
\ee
The usual geodesic equation for an affine parameter and without external potential is recovered 
using $N=\text{const.}$ and $V(x)=0$.

From the action (\ref{A01}) we can proceed to study the Hamiltonian formalism by defining the canonical momentum conjugate to $x^\mu$ as
\be
p_\mu \equiv \frac{\delta L}{\delta \dot{x}^\mu}  = \frac{m}{N}
g_{\mu\nu} \dot{x}^\nu \Rightarrow \dot{x}^\mu=\frac{N}{m}p^\mu,
\ee
and the corresponding momentum associated with the variable $N$:
\be
p_N\equiv\frac{\partial L}{\partial \dot{N}}=0.
\ee
Therefore, the variable $N$ will not be a dynamical variable.

The Hamiltonian is given by
\be\label{H01}
\mathcal{H}(p_\mu, x^\nu, \tau) \equiv p_\mu \dot{x}^\mu-L= N(x^\mu)\frac{m}{2}\!\left(\frac{p_\mu p^\mu}{m^2}+1+\!V(x) \right),
\ee
and the evolution of the constraint $p_N$ have to be constrained, so we can write another constraint as
\be\label{constraint}
C:=\dot{p}_N=\{p_N, \mathcal{H}\}=-\frac{m}{2}\left(\frac{p_\mu p^\mu}{m^2}+1+ V(x) \right)\approx 0,
\ee
where the symbol ``$\approx$'' stands for the evaluation of the constraint on solutions to the Hamilton equations. 
The evolution of the constraint $C$ is trivially zero on shell, so no further constraint has to be added to the formalism:
\be 
\dot{C}=\{C,\mathcal{H}\}=-\{C,N\}\,C\approx 0.
\ee

Making use of the constraint \eqref{constraint} the following relation for the cuadrimomentum is obtained:
\be
 p_\mu p^\mu\approx-m^2(1+V(x)),\label{HC01}
\ee
and the Hamiltonian can be finally written as $\mathcal{H}=-NC$.

The equations of motion described by the Hamiltonian \eqref{H01} will be given by
\beq
\dot{x}^\mu &=& \frac{N}{m}p^\mu,\label{Heq01}\\
\dot{p}_\mu &=& \{p_\mu, \mathcal{H}\}=-\{p_\mu, N\}C-N\{p_\mu, C\}\nonumber\\
&\approx& N \left(\frac{1}{m}g^{\beta \lambda} \Christoffel{g}^\kappa_{\mu\beta} p_\kappa p_\lambda-\frac{m}{2}\partial_\mu V(x)\right).\label{Heq02}
\eeq
These equations may be combined to obtain the equation of the geodesics, Eq. (\ref{GEO01}).

At this point, we will consider two equations of the kind of Eq. (\ref{GEO01}). The first one would correspond to a curve described by the metric $g_{\mu\nu}$, the parametrization $N$, and under the potential $V$. On the other hand, we will have the curve described by the metric $q_{\mu\nu}$, the parametrization $\tilde{N}$, and under the potential $\tilde{V}$:
\beq
\ddot{x}^\mu + \Christoffel{g}^\mu_{\alpha \beta} \dot{x}^\alpha \dot{x}^\beta = \frac{1}{N}\dot{x}^\mu \dot{x}^\nu
\partial_\nu N-\frac{N^2}{2} g^{\mu\nu}\partial_\nu V\,,\label{original}\\
\ddot{x}^\mu + \Christoffel{q}^\mu_{\alpha \beta} \dot{x}^\alpha \dot{x}^\beta = \frac{1}{\tilde{N}}\dot{x}^\mu \dot{x}^\nu
\partial_\nu \tilde {N}-\frac{\tilde{N}^2}{2} q^{\mu\nu}\partial_\nu \tilde{V}\,.\label{geodesicq}
\eeq
If we perform a conformal transformation of the metric ${g}_{\mu\nu}=\Omega^2 q_{\mu\nu}$, given that the Christoffel symbols of the two metrics are related by
\be 
 \Christoffel{q}^\mu_{\nu\rho}= \Christoffel{g}^\mu_{\nu\rho}
 -2\delta^\mu_{(\nu}\partial_{\rho)}\ln \Omega+g_{\nu\rho}g^{\mu\sigma}\partial_\sigma\ln \Omega,
\ee
and choosing the following relation between the function $\tilde{N}$ and the conformal factor $\Omega$, $\Omega^2=\frac{N}{\tilde{N}}$, the geodesic equation (\ref{geodesicq}) can be written in the following way:
\beq
\ddot{x}^\mu +  \Christoffel{g}^\mu_{\alpha \beta} \dot{x}^\alpha \dot{x}^\beta &=&-\frac{1}{2}N^2(1+V(x)){g}^{\mu\nu}\partial_\nu \ln\tilde{N}\nonumber\\ 
&&-\frac{N\tilde{N}}{2} g^{\mu\nu}\partial_\nu \tilde{V}\,.
\eeq
This equation would describe the same curve as equation \eqref{original} given that the right hand side of both equations coincide:
\beq
\hspace*{-0.8cm}\frac{1}{N}\dot{x}^\mu \dot{x}^\nu
\partial_\nu N-\frac{N^2}{2} g^{\mu\nu}\partial_\nu V\!\!\! &=&\!\!\!
-\frac{1}{2}N^2(1+V(x)){g}^{\mu\nu}\partial_\nu\! \ln\tilde{N}\nonumber\\
&&\!\!\!-\frac{N\tilde{N}}{2} g^{\mu\nu}\partial_\nu \tilde{V}\,.
\eeq
The choice of the parameter of the curve in \eqref{original} is encoded in the choice of the function $N$. If we choose $N=\text{const}.$ (an affine parameter), and the potential $V=0$, then the condition for the previous equation to describe the same curve is
\be 
\tilde{V}=\f{N}{\tilde{N}}=\Omega^2\,.
\ee
Therefore, it is possible to consider a non-affine parameter $\lambda$ defined by $\tilde{N}$ in the following way:
\be
-\tilde{N}^2d\tau^2=ds^2=-N^2d\lambda^2\Rightarrow \dot{\lambda}=\f{1}{\tilde{V}}\,.
\ee
In terms of this new parameter, the equation \eqref{geodesicq} takes the form:
\be 
\f{d^2x^\mu}{d\lambda^2}+\Christoffel{q}^\mu_{\alpha\beta}\f{dx^\alpha}{d\lambda}\f{dx^\beta}{d\lambda}=-\f{N^2}{2}q^{\mu\nu}\partial_\nu\tilde{V}\,.\label{geodesicqlambda}
\ee

Summarizing, we have deduced that it is possible to describe a geodesic of a certain metric $g_{\mu\nu}$ that is described using an affine parameter as a non-geodesic curve described by a non-affine parameter (encoded by $\tilde{N}\neq \text{const}.$) of a conformal metric $q_{\mu\nu}=g_{\mu\nu} \tilde{N}/N$ under the action of a non-trivial potential $\tilde{V}=N/\tilde{N}$.

\subsection{Recovering Newtonian mechanics}\label{Newtoniansection}

We will explore now the small energy limit (with respect to the mass at rest of the test particle) or the small velocity limit (with respect to the speed of light) of the theory of relativity. As it is well known, it  gives rise to the Newtonian description of the movement of a particle. In order to show it, let us restrict to spacetimes described, in the appropriate limit, by a static metric (the slow varying gravitational field approximation)
\be
ds^2 = g_{\mu\nu} dx^\mu dx^\nu =  g_{tt} dt^2 + g_{ij} dx^i dx^j .
\ee
Let us consider a conformal metric $q_{\mu\nu}$ defined by
\be
q_{\mu\nu}=\frac{g_{\mu\nu}}{|g_{tt}|}.
\ee
In order to explore the Newtonian limit of the metric $g_{\mu\nu}$, we will consider that  $g_{\mu\nu}\approx \eta_{\mu\nu}+h_{\mu\nu}$, with $|h_{\mu\nu}|\ll 1$.

The action \eqref{A01} for the new metric, parametrization and potential is
\be
S[x^\mu(\tau)] \!
=\! \frac{m}{2}\!\int\!\!\left(\frac{1}{\tilde{N}^2} q_{\mu\nu} \dot{x}^\mu \dot{x}^\nu-(1+ \tilde{V}(x)) \right) \tilde{N} d\tau ,
\ee
that, written in terms of the non-affine parameter $d\lambda\equiv\tilde{N} d\tau $ takes the form:
\be
S[x^\mu(\tau)] \!
=\! \frac{m}{2}\!\int\!\!\left(q_{\mu\nu} \f{dx^\mu}{d\lambda}\f{dx^\nu}{d\lambda}-(1+ \tilde{V}(x)) \right) d\lambda\,.
\ee
In the slow motion approximation, we have that:
\be
-\tilde{N}^2d\tau^2=-d\lambda^2\sim -dt^2(c^2-v^2)=-c^2dt^2(1-\epsilon^2)\,,
\ee
with $v^2=\f{dx^i}{dt}\delta_{ij}\f{dx^j}{dt}$ and $\epsilon=v/c\ll 1$.
At first order in $\epsilon$ the following relation is satisfied: 
\be
\tilde{N}d\tau=dt\Rightarrow \dot{t}=\tilde{N}\Rightarrow \f{dt}{d\lambda}=1\,.
\ee
Then, the action takes the form:
\be
S[x^\mu(\tau)]=\frac{m}{2}\int\left(q_{\mu\nu} \frac{dx^\mu}{dt}\frac{dx^\nu}{dt}-(1+\tilde{V}(x)) \right) dt .
\ee
In this case $\tilde{V}(x)=\Omega^2=|g_{tt}|=(1+h_{tt})$, and
\be
q_{\mu\nu} \frac{dx^\mu}{dt}\frac{dx^\nu}{dt}=-1+q_{ij}v^iv^j\,.
\ee
Given that at first order $q_{ij}v^iv^j\sim v^2$, we obtain, up to total derivatives with no effect on the variational principle, 
\be
S[x^\mu(\tau)]=
\int\left(\frac{1}{2}mv^2-m\frac{h_{tt}}{2} \right) dt.\label{A04}
\ee
The action \eqref{A04} is the Newtonian action of a test particle moving under the action of the potential $\phi(x)=h_{tt}/2$ in a flat three dimensional space.
The temporal component of the original metric is, as expected from the Newtonian limit of general relativity:
\be 
g_{tt}=-(1+2\phi).
\ee

From the Hamiltonian point of view, the Hamiltonian corresponding to the conformal metric is:
\be
\tilde{\mathcal{H}}= \tilde{N}(x^\mu)\frac{m}{2}\left(\frac{p_\mu p_\nu q^{\mu\nu}}{m^2}+1+\tilde{V}(x) \right)=-\tilde{N}\tilde{C}.
\ee
To compute the evolution of a dynamical variable $f$ we proceed as follows:
\be
\dot{f}\equiv\f{df}{d\tau}=-\{f,\tilde{N}\}\tilde{C}-\tilde{N}\{f,\tilde{C}\}\approx-\tilde{N}\{f,\tilde{C}\},
\ee
that, in terms of the non-affine parameter $\lambda$ will be given by:
\be 
\frac{df}{d\lambda}=-\{f,\tilde{C}\}\,.
\ee
Therefore, in the Newtonian limit ($dt/d\lambda=1$)
\be 
\frac{df}{dt}=-\{f,\tilde{C}\}\,,\label{evolutiont}
\ee
and, taking $t$ as our time for dynamical evolution, the Hamiltonian is written as
\be 
H_{\text{Newt.}}=\frac{1}{2 m }{q^{ij}} p_i p_j+m\f{\tilde{V}}{2},\label{HNewt1}
\ee
with $p_i=mq_{ij}\f{dx^j}{dt}$. This Hamiltonian \eqref{HNewt1} is the Newtonian Hamiltonian for  a particle in the 3-geometry given by the spatial part of the conformal metric $q_{ij}$ under the action of a potential $\phi=\tilde{V}/2$.

\subsection{De Sitter spacetime in static coordinates}\label{deSitter}

As an example, let us consider the de-Sitter spacetime, described in static coordinates by the line element
\be
ds^2 = g_{\mu\nu} dx^\mu dx^\nu =  - \Delta dt^2 + \Delta^{-1} dr^2  +r^2( d\theta^2+\sin^2\theta d\vphi) \label{DSM01} ,
\ee
where $\Delta = (1-\frac{\Lambda r^2}{3})$, in units for which $c=G=1$.

The conformal metric $q_{\mu\nu}=g_{\mu\nu}/|g_{tt}|$ turns out to be
\be
q_{\mu\nu} = {\rm diag}\left( - 1, \Delta^{-2}, \Delta^{-1} r^2 , \Delta^{-1} r^2 \sin^2\theta \right).
\ee
Therefore, the potential $\tilde{V}$ is a central potential given by:
\be
\tilde{V}=\Omega^2=|g_{tt}|=\Delta.\label{potentialDeSitter}
\ee

In this case, computing the equation \eqref{geodesicqlambda} for the time coordinate, we obtain that $d^2t/d\lambda^2=0$, because of $\Christoffel{q}^t_{\alpha \beta}=0$. So we can set, in this precise example of de Sitter, that $dt/d\lambda=1$. Notice that, in this case, this is true without the necessity of imposing any approximation (for a general metric this would be only true in the Newtonian limit, as commented in the previous section).

Therefore, for the case of de Sitter, the evolution of any dynamical variable $f$ will be exactly given by the equation \eqref{evolutiont} with the three dimensional Hamiltonian given by:
\be 
H_{\text{3d}}=\frac{1}{2 m }{q^{ij}} p_i p_j+m\frac{\Delta}{2},\label{H3d}
\ee
with $p_i=mq_{ij}\f{dx^j}{dt}$. 
This Hamiltonian \eqref{H3d} is the generator of the evolution
of a particle in the 3-geometry given by the spatial part of the conformal metric $q_{ij}$ under the action of a potential $\phi=\Delta/2$. In this way, the free falling of a test particle in de-Sitter spacetime can be seen as a non geodesic trajectory of a spatial geometry $q_{ij}$ under the potential $\phi$. Then, an observer located at the origin of coordinates would see the movement of a distant galaxy that is still far from the corresponding event horizon as following the action of the Hamiltonian \eqref{H3d}.

Now we can explore the Newtonian limit in this system, that is fulfilled if $g_{\mu\nu}\approx \eta_{\mu\nu}+h_{\mu\nu}$, with $|h_{\mu\nu}|\ll 1$. It will be valid if
\be
|h_{tt}|=\frac{\Lambda r^2}{3}\ll 1,
\ee 
that is, in a distant region from the event horizon ($\Delta \approx 1$) approximately describes the flat space
\be
ds^2 = -dt^2+dr^2 + r^2 d\theta^2 + r^2 \sin^2\theta d\varphi^2 .
\ee
In particular, within the Newtonian approximation, the non geodesic equation (\ref{geodesicqlambda}) for the variable $r$ of a radial trajectory would approximately be the Newton second law
\be
\frac{d^2r}{dt^2} \approx - \partial_r (\tilde{V}/2) = \frac{\Lambda}{3} r\,,
\ee 
which, considering that $r(t) = a(t) r_0$, is equivalent to the second Friedmann equation,
\be
\frac{\ddot a}{a} = \frac{\Lambda}{3} .
\ee
Of course, near the cosmological horizon the Newtonian approximation breaks down, the Newtonian potential vanishes, $\tilde{V} \approx 0$, and the trajectory of the test particle, computed directly with the equation \eqref{geodesicqlambda} becomes frozen:
\be
r(t) \approx \sqrt{\frac{3}{\Lambda}} \tanh \sqrt{\frac{\Lambda}{3}} (t-t_0)\, \xrightarrow{t\rightarrow \infty}\,\sqrt{\frac{3}{\Lambda}} ,
\ee
showing that the observer cannot see the test particle crossing the event horizon.


\section{Quantisation of spacetime coordinates}

In the previous section we have dealt with the possibility of describing the geodesics of a certain metric as the non-affine geodesics of other metric conformally related to the previous one under a potential given by the conformal factor used. Therefore, up to now, all the treatment has been classical.

Nevertheless, the procedure described in section \ref{Lagrangian} is useful to study the canonical quantisation of the system and to recover the Schr\"odinger equation starting from a metric. In this way, we are able to associate to a metric written in certain coordinates, a Schr\"odinger equation subject to a potential.

\subsection{Canonical quantisation of spacetime coordinates}\label{Quant}

The formalism described in section \ref{Lagrangian} allows us to apply a canonical quantisation of the spacetime coordinates. It is important to notice that the time coordinate $t$ has been treated on the same footing of any other coordinate so it is going to be quantised in the same way as the spatial coordinates. Thus, let us start by promoting the spacetime variables and their conjugate momenta to operators,
\be\label{OP01}
x^\mu \rightarrow \hat{x}^\mu \, , \, p_\mu \rightarrow \hat{p}_\mu .
\ee
In the Schr\"{o}dinger picture the operators $\hat x^\mu$ and $\hat p_\mu$ are fixed and the evolution of the quantum state of the coordinates of a test particle is completely encoded in a wave function that depends on the variables of the configuration space, i.e. the spacetime coordinates, $\phi(x)$. The action of the operators (\ref{OP01}) onto the wave function $\phi(x)$ is
\be\label{CQ01}
\hat{x}^\mu  \phi(x) = x^\mu  \phi(x) \, , \, \hat{p}_\mu \phi(x) = - i\hbar \frac{\partial \phi(x)}{\partial x^\mu} .
\ee
The Hamiltonian constraint (\ref{HC01}), with $V(x)=0$, turns out to be then
\be\label{WE01}
\left( - \hbar^2 \Box_x + m^2 \right) \phi(x) = 0 ,
\ee
where, with an appropriate choice of factor ordering,
\be
\Box_x \equiv \frac{1}{\sqrt{-g}} \partial_\mu\left( \sqrt{-g} g^{\mu\nu} \partial_\nu\right) .
\ee
Equation (\ref{WE01}) is equivalent, with a rescaling of the mass $m \rightarrow \hbar m$, to the Klein-Gordon equation of a scalar field that propagates in a curved spacetime, which can be obtained from the customary Lagrangian formalism of a scalar field (see Eqs. (3.24) and (3.26) of Ref. \cite{Birrell1982}, with a different sign however with respect to (\ref{WE01}) due to the different signature chosen here).  
However, it is worth noticing that (\ref{WE01}) has been obtained by merely quantising the coordinates of the spacetime in the classical Hamiltonian constraint (\ref{HC01}) with a consistent choice of factor ordering. 
Nevertheless, notice that in our case, $\phi(x)$ is not a classical field (as it is when considering the usual Klein-Gordon equation), but a quantum wave function whose evolution is given by \eqref{WE01}.

\subsection{Geodesic equation from the quantum state}\label{geodesic}

As it has been pointed out in the previous section, the wave equation (\ref{WE01}) is not a classical equation but a quantum one, which is clear from the existence of the factor $\hbar^2$ in the kinetic term. In particular, that factor allows us to make an expansion of the wave equation (\ref{WE01}) in powers of $\hbar$. The term $\hbar^0$ of that expansion must revert to the classical Hamiltonian constraint (\ref{H01}). 

Thus, the wave function $\phi(x)$ that appears in the Klein-Gordon equation (\ref{WE01}) contains not only information about the matter field that propagates along the spacetime but it has also information about the spacetime itself through the information contained in the geodesics, which encapsulate the geometrical structure of the given spacetime. In order to show it, let us consider the WKB solutions of the equation (\ref{WE01}),
\be\label{WKB01}
\phi(x) = C(x) e^{\pm \frac{i}{\hbar} S(x)} ,
\ee
where $C(x)$ is a slow varying function of the spacetime coordinates and $S(x)$ is the action of the spacetime. Inserting the wave function (\ref{WKB01}) into the Klein-Gordon equation (\ref{WE01}) we obtain, at zeroth order in $\hbar$, 
\be\label{HJ01}
g^{\mu\nu} \frac{\partial S}{\partial x^\mu} \frac{\partial S}{\partial x^\nu} + m^2 = 0 ,
\ee
which is the Hamilton-Jacobi equation of the coordinates of the  test particle. One can compare this equation (\ref{HJ01}) with the Hamiltonian constraint (\ref{HC01}) and identify the corresponding momenta as
\be
p_\mu = \frac{\partial S}{\partial x^\mu} ,
\ee
and therefore,
\be\label{HJc01}
\dot{x}^\mu = \pm \f{N}{m}g^{\mu\nu} \frac{\partial S}{\partial x^\nu} = \pm\f{N}{m} g^{\mu\nu} p_\nu ,
\ee
where we have defined a WKB parameter $\tau$, through the relation
\be\label{WKBtau}
\frac{\partial}{\partial \tau} = \pm\f{N}{m} g^{\mu \nu} \frac{\partial S}{\partial x^\mu} \frac{\partial }{\partial x^\nu} .
\ee
The two signs in (\ref{HJc01}) and (\ref{WKBtau}), which correspond to the two different solutions given in (\ref{WKB01}), make explicit the invariance of the action (\ref{ACT01}) with respect to the reversal of the affine parameter. It means that we have two symmetric solutions, each one representing the same curve but being run in opposite directions of their tangent vectors. We know from quantum mechanics that these two symmetric solutions correspond to a particle and an antiparticle moving in opposite time directions. 

With the use of the relation (\ref{HJc01}) the Hamilton-Jacobi equation (\ref{HJ01}) turns out to be
\be
\f{1}{N^2} g_{\mu\nu} \dot x^\mu \dot x^\nu + 1 = 0 ,
\ee
which together with (\ref{HJc01}) gives rise to the equations (\ref{Heq01}-\ref{Heq02}) and the corresponding geodesic equation (\ref{GEO01}), with $V(x)=0$. 
Notice also that in terms of the proper time $ds$, with the value $N=m$, the following expression is satisfied:
\be
ds^2 = g_{\mu\nu} dx^\mu dx^\nu = - m^2 d\tau^2 ,
\ee
where, by definition, $d\tau^2 >0$. Therefore, the value of the constant $m$, which in the field theory is the mass of the test particle, determines the character of the geodesic followed by the particle. For $m=0$ the trajectory of the particle is a null geodesic, as it should be, and for $m^2>0,$ it is a time-like geodesic.

Therefore, the wave function $\phi(x)$ already contains at the classical level (i.e., at zeroth order in $\hbar$) the geometrical structure of the spacetime through the equation of the geodesics. Of course, it also contains information about the matter fields that propagate in the background spacetime that can be obtained as usual from the quantum field theory of the Klein-Gordon equation (\ref{WE01}).

\subsection{Geodesics of de Sitter spacetime from the Bunch-Davies vacuum}\label{BDsection}

As an example, we will obtain the equations of the geodesics of a flat de Sitter spacetime from the well known solutions of the corresponding Klein-Gordon equation, the so called Bunch-Davies vacuum. Let us consider the half of the de Sitter spacetime covered by the flat coordinates, with metric element given by
\be\label{metricDeSitter}
ds^2 = - dt^2 + a^2(t) \left(dr^2 + r^2(d\theta^2+\sin^2\theta d\varphi^2)\right) ,
\ee
where the scale factor satisfies, $\dot a(t) = H a(t)$, with $H\equiv \frac{\Lambda}{3}$ and $\Lambda$ the cosmological constant. In the present formalism, if we choose a parameter such that $N=1$ for the description of the geodescics of the metric \eqref{metricDeSitter}, the Lagrangian of the action (\ref{A01}) reads
\be
L = \frac{m}{2} \left( - \dot t^2 + a^2 \dot r^2 - 1\right) ,
\ee
where for simplicity we have omitted the angular variables (we shall restrict the movement of the particle to the values $\dot \theta =  \dot\varphi = 0$). The equation of the geodesics can be easily  obtained from the corresponding Euler-Lagrange equations,
\beq\label{dott}
\ddot t + H a^2 \dot r^2 &=& 0 , \\ \label{pr}
p_r \equiv ma^2 \dot r &=& k_0 ,
\eeq
where we have used, $\dot a=H a$, and $k_0$ is a constant. The Hamiltonian constraint \eqref{HC01} for this case, and written with the configuration variables, will be:
\be
-\dot t^2 + a^2 \dot r^2=-1\,.\label{HdS01}
\ee
From (\ref{pr}) and (\ref{HdS01}) one obtains
\be\label{OM01}
\dot t = \pm \f{1}{m}\sqrt{\frac{k_0^2}{a^2} + {m^2}} \equiv \pm \f{\omega(t)}{m} .
\ee
where $\omega(t)$ has been defined here for later convenience. With the use of $a(t)\propto e^{Ht}$ this equation could be integrated to obtain the path of the geodesic $t(\tau)$.

On the other hand, the Klein-Gordon equation (\ref{WE01}) for a flat de Sitter spacetime reads\footnote{Omitting the angular terms.} 
\be\label{WEdS01}
\frac{\hbar^2}{a^3} \partial_t\left( a^3 \partial_t \phi \right) - \frac{\hbar^2}{a^2 r^2} \partial_r\left( r^2 \partial_r \phi \right) + m^2 \phi = 0 ,
\ee
and its solutions are well-known. Using the ansatz\footnote{
Notice that, given that we are not considering the angular variables, plane waves instead of spherical harmonics appear.} 
\be\label{Sep01}
\phi(t,r) = \frac{1}{r} e^{\pm  \frac{i}{\hbar} k_0 r} \phi_t(t) ,
\ee
it is obtained the customary equation for the temporal component $\phi_t(t)$,
\be\label{KGdS02}
\hbar^2 \ddot \phi_t(t)  + 3 \hbar^2 \frac{\dot a}{a} \dot \phi_t(t) + \omega^2 \phi_t(t) = 0 ,
\ee
where $\omega = \omega(t)$ is defined by equation (\ref{OM01}). Notice the appearance of the $\hbar^2$ factor in the  derivative terms of (\ref{KGdS02}). This fact will allow us to expand the solutions in powers of $\hbar$ and, in particular, it will allow us to show that the classical equations of the geodesics of the de Sitter spacetime can be obtained from the $\hbar^0$ order expansion of the solution of (\ref{WEdS01}).

\begin{figure}[t]
\begin{center}
       \includegraphics[width=0.45\textwidth]{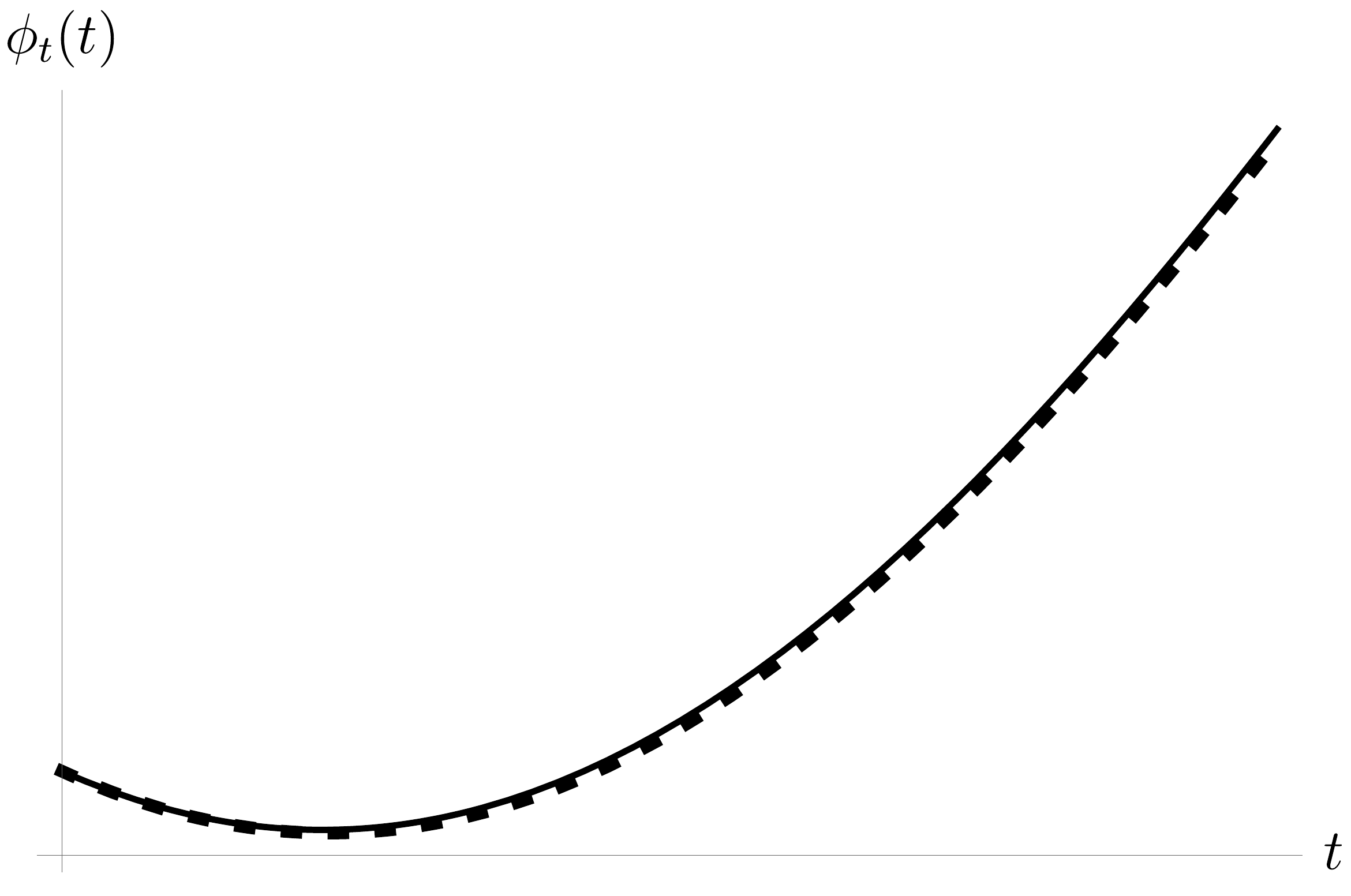}
\end{center}
       \caption{This plot shows, for a set of values of the parameters within the regime of interest (large modes, $\hbar\rightarrow 0$), the real part of the solution to equation \eqref{KGdS02} given by the Bunch-Davies vacuum \eqref{BD01} --dashed line-- and the real part of its WKB approximation \eqref{WKBdS01} --continuous line--. The complete overlapping of the solution and its approximation is observed.}
    \label{figure_bunch}
\end{figure}

In order to solve (\ref{KGdS02}) it is customary to make the change, $\chi = a \phi$, and to use conformal time, $\eta = \int\frac{dt}{a}$, in terms of which it becomes
\be
\chi'' + \left( k^2 + m^2 a^2 - \frac{a''}{a^2} \right) \chi = 0 ,
\label{chi}
\ee
where $f'\equiv\f{df}{d\eta}$. Solutions to equation \eqref{chi}
are well known in terms of Bessel or Hankel functions \cite{Birrell1982, Mukhanov2007}. Unravelling the changes, we obtain
\be\label{BD01}
\phi_{BD}(t,r) =  \sqrt{\frac{\pi}{H a^3}} \frac{1}{2 r} e^{  \frac{i}{\hbar} k_0 r}  \mathcal{H}_n^{(2)}\left( \frac{k_0}{\hbar H a} \right) ,
\ee
where $\mathcal{H}_n^{(2)}$ is the Hankel function of the second kind and 
\be\label{BDn}
n = \sqrt{\frac{9}{4}- \frac{m^2}{\hbar^2 H^2}} .
\ee
Notice the appearance of the Planck constant $\hbar$  in (\ref{BD01}) and (\ref{BDn}) in contrast to the customary solutions of the Bunch-Davies vacuum (see, for instance, Refs. \cite{Birrell1982, Mukhanov2007}). It allows us to expand the wave function (\ref{BD01}) in powers of $\hbar$. In order to explore the semiclassical limit $\hbar\rightarrow 0$, let us use the Debye asymptotic expansion for the Hankel functions \cite{Matviyenko1992},
\be
\mathcal{H}^{(2)}_\nu(x) = \sqrt{ \frac{2}{\pi}}  (x^2 - \nu^2)^{-\frac{1}{4}} e^{- i \eta_1} \left( 1 + \mathcal O(\frac{1}{\nu})\right) ,
\ee
with,
\be
\eta_1 = (x^2 - \nu^2)^\frac{1}{2} - \nu \, \arccos(\frac{\nu}{x}) -\frac{\pi}{4} .
\ee
Then, using the values
\be
x = \frac{k_0}{\hbar H a}, \quad \ -i \nu = \alpha = \frac{m}{\hbar H} ,
\ee
and the fact that
\be
- i \alpha \arccos\frac{i \alpha}{x} = - i\frac{\pi \alpha}{2} - \alpha \log\left( \frac{\alpha}{x} + \sqrt{\frac{\alpha^2}{x^2}+1}  \right),
\ee
the following approximation of the Hankel function of the second kind is obtained
\be
\mathcal{H}^{(2)}_{\frac{i m}{\hbar H}}\left( \frac{k_0}{\hbar H a} \right) \approx \sqrt{\frac{2\hbar H}{\pi \omega}} e^{-\frac{\pi m}{2 \hbar H}} e^{-\frac{i}{\hbar}\left( \frac{\omega}{H} - \frac{m}{ H} \log\left(\frac{a}{k_0}(m+\omega) \right) \right) } ,
\ee
where $\omega(a)$ is given in (\ref{OM01}). Thus, we arrive at the following WKB approximation of the Bunch-Davies wave function 
\be\label{WKBdS01}
\phi_W(t,r) = \frac{N_0}{r\sqrt{a^3(t) \omega(t)}} e^{\frac{i}{\hbar} S(t,r) } ,
\ee
where $N_0$ is a constant, and
\be\label{SWKB}
S(t,r) = k_0 \, r - \frac{\omega}{H} + \frac{m}{ H} \log\left(\frac{a}{k_0}(m+\omega) \right)  .
\ee
The Debye asymptotic expansion is guaranteed for real values of the argument (see, Ref. \cite{Olver1974, Matviyenko1992} and references therein). Even though in our case $\nu$ is complex, we can check that, for the range of the parameters we are interested in, it is still a good approximation.
In figure \ref{figure_bunch} we can observe that the solution to the equation \eqref{KGdS02} given by the Bunch-Davies vacuum \eqref{BD01} and the WKB approximation \eqref{WKBdS01} are completely overlapped. 
In this sense, we expect the WKB solution to be a good approximation for large values of the argument of the Hankel function, that is, for $\f{k_0}{\hbar H a}\rightarrow \infty$. Therefore, we would expect the approximation to be valid for large values of the mode $k_0$, that means wavelengths much smaller than the cosmological horizon. If the wavelength considered were of the order of the cosmological horizon the quantum corrections would not be negligible anymore and they should be taken into account.

Inserting the wave function (\ref{WKBdS01}) into the Klein-Gordon equation (\ref{WEdS01}), it is obtained at $\hbar^0$ order,
\be
-\left( \frac{\partial S}{\partial t} \right)^2 + \frac{1}{a^2} \left( \frac{\partial S}{\partial r} \right)^2 + m^2 = 0 ,
\ee
which is directly satisfied because, from (\ref{SWKB}),
\be\label{prpt}
\frac{\partial S}{\partial r}  = k_0  \ , \ 
\frac{\partial S}{\partial t} = \omega(t)  ,
\ee
with $\omega(t)$ given by (\ref{OM01}). Furthermore, from (\ref{pr}) and (\ref{OM01}), we see that these values correspond to the momenta conjugated to $r$ and $t$, $p_r = k_0$ and $p_t = - \omega$, respectively. 

We can now define an affine parameter $\tau$ by the relation
\be
\frac{\partial}{\partial \tau} \equiv \pm \f{1}{m}\left( \omega(t) \frac{\partial}{\partial t} + \frac{k_0}{a^2} \frac{\partial }{\partial r} \right) ,
\ee
so we obtain 
\be
\dot t = \pm \f{\omega}{m} \ , \ \dot r = \pm \frac{k_0}{m a^2} ,
\ee
which can be combined to yield the geodesic equations (\ref{dott}) and (\ref{pr}). Therefore, the equation of the geodesics of the de Sitter spacetime can be obtained from the quantum state of the corresponding Klein-Gordon equation. This result is remarkable because it  shows that the quantum state of a field that propagates along a given spacetime contains not only  information about the  state of the field but also  about the geometrical structure of the spacetime where it propagates through, because of the information contained in the geodesic equations. 
Besides, it will allow us to analyse quantum corrections to a geodesic path by considering not only the $\hbar^0$ terms but also $\hbar^1$ and higher powers in $\hbar$, whose effect is expected to be significant when the path approaches a singular region of the spacetime. Thus, the analysis might hopefully end up in the avoidance of the classical singularities due to the quantum corrections of the geodesics \cite{RP2018d}.

\subsection{Recovering Schr\"{o}dinger equation}\label{Sch}

For completenes, we shall now show that the Klein-Gordon equation (\ref{WE01}) can also be reduced in the non relativistic limit to the Schr\"{o}dinger equation. Thus, the quantisation of the spacetime coordinates relates the relativistic description of the spacetime and the quantum description of matter, providing a framework were it is possible to join the two fundamental, in principle irreconcilable, theories of the physics of the XXth century. 

Let us first focus on spacetimes whose line element can be written in the appropriate limit as
\be
ds^2=g_{\mu\nu}dx^\mu dx^\nu = - dt^2 + g_{i j} dx^i dx^j .
\ee
In that case, the Hamiltonian constraint (\ref{HC01}) with the potential $V=0$, turns out to be
\be
- p_t^2 + g^{i j} p_i p_j + m^2 = 0 .
\ee 
Therefore, the quantum version of the Hamiltonian constraint, given by (\ref{WE01}), becomes
\be
\hbar^2  \frac{\partial^2 \phi}{\partial t^2} - \hbar^2 \nabla^2\phi + m^2 \phi = 0 ,
\ee
where the Laplacian $\nabla^2$ operator is given by
\be
\nabla^2 = \frac{1}{\sqrt{-g}} \partial_i \left( \sqrt{-g} \, g^{i j} \partial_j\right) .
\ee
We can now consider semiclassical solutions of the form
\be\label{WKB03}
\phi(x) = \frac{1}{\sqrt{2 m }} e^{- \frac{i}{\hbar} S(t) } \psi(\vec x, t) ,
\ee
where 
\be\label{mt01}
S(t) = \int dt \, m = m (t-t_0)  .
\ee
The wave function (\ref{WKB03}) is valid for regions of the spacetime for which the variation of the background time variable $t$ is small compared with respect to the variation of the spatial variables, i.e., in the non relativistic regime for velocities $v\ll c$. In this regime, $\dot{t}\approx 1$, so $p_t \approx -m$.

It means that the kinetic term that corresponds to the spatial velocities is small compared with the mass at rest of the test particle. In the semiclassical regime, the Klein-Gordon equation can be expanded in powers of $\hbar$ and disregard terms of order $\hbar^2$ in the time variable. Then, it is obtained
\be
2 i \hbar  \frac{\partial S}{\partial t}  \frac{\partial \psi}{\partial t} = - \hbar^2 \nabla^2 \psi ,
\ee 
which, using (\ref{mt01}), can be written as
\be
i \hbar   \frac{\partial \psi}{\partial t} = - \frac{\hbar^2}{2 m} \nabla^2 \psi ,
\ee
which is the Schr\"{o}dinger equation of a free particle propagating in the space described by the metric $h_{ij}$.

In the case of a spacetime with line element given by
\be
ds^2 = - |g_{tt}| dt^2 + g_{i j} dx^i dx^j ,
\ee
one can use the development of Sect. II to change to a conformal metric $q_{\mu\nu}=\f{1}{|g_{tt}|}g_{\mu\nu}$ using a potential $\tilde{V}(x)=|g_{tt}|$ and a parametrization $\tilde{N}=1/|g_{tt}|$. In this way, using equation \eqref{HC01}, we arrive at the classical Hamiltonian constraint
\be\label{HC03}
q^{\mu\nu} \tilde{p}_\mu \tilde{p}_\nu = -m^2(1+|g_{tt}|),
\ee
where
\be
\tilde p_\mu = \f{m}{\tilde{N}} q_{\mu\nu} \dot{x}^\nu.
\ee
The Hamiltonian constraint (\ref{HC03}) turns out to be then
\be
-\tilde p_t^2 + q^{ij} \tilde{p}_i \tilde p_j +  m^2(1+ \tilde{V}(x)) = 0 ,
\ee
which under canonical quantisation of the new variables $(x^\mu, \tilde p_\mu)$ transforms into
\be
\hbar^2  \frac{\partial^2 \phi}{\partial t^2} - \hbar^2 \tilde \nabla^2\phi + m^2(1+ \tilde{V}(x))  \phi = 0 ,
\ee
where the Laplacian $\tilde \nabla^2$ operator is now given by
\be
\tilde \nabla^2 = \frac{1}{\sqrt{-q}} \partial_i \left( \sqrt{-q} \, q^{i j} \partial_j\right) .
\ee
A similar wave function to (\ref{WKB03}) can now be used to obtain 
\be\label{SCEQ01}
i \hbar   \frac{\partial \psi}{\partial t} = \left( - \frac{\hbar^2}{2 m} \tilde \nabla^2 + \f{m}{2} \tilde V(x) \right) \psi ,
\ee
Equation (\ref{SCEQ01}) is the Schr\"{o}dinger equation of a particle of mass $m$ moving under the action of the potential $\tilde{V}(x)$ in the space geometrically described by the metric tensor $q_{ij}$ and has been derived from the Klein-Gordon equation, which in turn has been obtained from the canonical quantisation of the spacetime coordinates of a given spacetime. Thus, general relativity and quantum field theory are related by the quantisation of the spacetime coordinates and so they are their non relativistic limits, the Newtonian limit and the Schr\"{o}dinger description of quantum mechanics.


\section{Conclusions}\label{conclusions}

In spite of the attempts to give a quantum description of the gravitational interaction it remains as one of the main open problems in Theoretical Physics.
In the present paper we worked out a proposal for canonically quantise the coordinates of the spacetime in a reference system attached to an observer within the framework given by the general relativity theory. In this sense, we propose a quantisation of the spacetime suitable for a certain observer.
Although it is clearly an observer dependent quantisation, it should provide a good  approximation to a possible full background independent quantisation of the geometry of the spacetime.

Firstly, we explored the classical description of the geo\-desics. Using the action principle we obtained the equation of the geodesics of a given metric.
Interestingly, we realized that  these same curves could be described as a non-geodesic equation parameterized by a non-affine parameter of a different metric related with the first one by a conformal transformation and under a certain potential written in terms of the conformal factor used.
Furthermore, the Hamiltonian formalism with the constraints of the system was also studied.

The Newtonian limit was also explored within this formalism. As expected, the Newton equation appears as the low energy limit of general relativity. In addition, using the perspective given by the non-affine geodesics under a certain potential we were able to study the case of de Sitter spacetime in static coordinates. For this special case, an exact and equivalent Hamiltonian for a curved three dimensional space that resembles the Newtonian Hamiltonian is obtained. Afterwards, the Newtonian limit of this important cosmological scenario was also studied.

The canonical quantisation of the spacetime coordinates leads to a Klein-Gordon like equation that appears after writing the Hamiltonian constraint of the classical formalism \textit{\`a la} Wheeler-deWitt. Nevertheless, this Klein-Gordon equation corresponds to an equation for a quantum wave function, in contrast with the usual Klein-Gordon equation as the evolution of a classical scalar field. In fact, the appearance of a factor $\hbar$ in the coefficients of the equation remarks this issue.

Within this framework we are able to recover the classical geodesics corresponding to our quantum description of the spacetime starting from the quantum Klein-Gordon equation previously discussed. An example has been given in order to obtain the geodesic equation corresponding to the de Sitter spacetime taking as the starting point the Bunch-Davies vacuum of the quantum theory.

Finally, the recovery of the Schr\"odinger equation in the appropriate non-relativistic limit was discussed. We obtained the Schr\"odinger equation with a potential given by the temporal component of the metric, as it was expected from the fact that this is the Newtonian potential for the gravitational field.

Summarizing, in this article we managed to put together several aspects of the quantum theory that, a priori, remained separate and that showed incompatibilities at the fundamental level. Within the framework proposed here it is possible to recover from the quantum prescription of the geometry both the geodesics given by general relativity and their Newtonian limit, as well as the Schr\"odinger equation for the non-relativistic limit. This could shed light to the problem of full background independent quantisation of the spacetime and may be helpful to study certain simple models as the de Sitter universe presented in this paper.


\section*{Acknowledgements}

IG acknowledge financial support from project FIS2017-85076-P (MINECO/AEI/FEDER, UE), and Basque Government Grant No.~IT956-16.


\section*{References}

\bibliographystyle{elsarticle-num.bst}
\biboptions{sort&compress}

\end{document}